\documentclass{osa-article}
\journal{osac}
\articletype{Research Article}
\usepackage{lineno}

\begin{document}

\title{\title{Design, simulation and characterization of integrated photonic spectrographs for Astronomy II: Low-aberration Generation-II AWG devices with three stigmatic points}}

\author{Andreas Stoll,\authormark{1,*} Kalaga Madhav,\authormark{1} and Martin Roth\authormark{1}}

\address{\authormark{1}Leibniz-Institut f\"ur Astrophysik (AIP), An der Sternwarte 16, 14482 Potsdam, Germany}

\email{\authormark{*}astoll@aip.de} %% email address is required



\begin{abstract}
In the second part of our series on integrated photonic spectrographs for astronomy, we present theoretical and experimental results on the design, simulation and characterization of custom-manufactured silica-on-silicon Arrayed 
Waveguide Gratings (AWGs) constructed using the three-stigmatic-point method. We derive several mid-to-high resolution field-flattened AWG designs, targeting resolving powers of 11,000 - 35,000 
in the astronomical H-band, by iterative computation of differential coefficients of the optical path function. We use numerical simulations to study the imaging properties of the designs in a wide wavelength range between 1500 nm and 
1680 nm. We theoretically discuss the design-specific degradation of spectral resolving power at far-off-centre wavelengths and suggest possible solutions. In the experimental section, we provide characterization results of 
seven manufactured AWG devices of varying free spectral range and resolution. We obtain estimates on spectral resolving powers of up to 27,600 for polarized input at 1550 nm from measurements of the
channel transmission bandwidth. Furthermore, we numerically predict expected resolving powers of up to 36,000 in polarized mode and up to 24,000 in unpolarized mode for direct continuous imaging of the spectrum.
\end{abstract}

%%%%%%%%%%%%%%%%%%%%%%%%%%  body  %%%%%%%%%%%%%%%%%%%%%%%%%%
\section{Introduction}
Integrated photonic spectrographs on the basis of Arrayed Waveguide Gratings are a subject of active development driven mainly by astronomy and astrophotonics \cite{Cvetojevic:09, Cvetojevic:12, Gatkine:18, Gatkine:20}, 
whose special requirements call for customized AWG designs for e.g. high-resolution spectroscopy in the near infrared. Integrated photonics plays an increasingly important role in the miniaturization and mass reduction of spectroscopic 
and sensing instruments, which is especially beneficial for space-exploration oriented applications. Arrayed Waveguide Gratings, originally introduced as wavelength demultiplexers in telecommunications, are modified for spectroscopic 
use by removal of the output waveguide array and polishing of the image plane facet to optical quality. The spectral image is captured by a near infrared (NIR) camera on a flat image detector. Canonical AWG designs based on the Rowland geometry 
known from telecommunications were studied in a previous work \cite{Stoll1:21}, hereafter paper I, with the intent to determine the performance of various AWG designs in the resolving power range between $15,000$ and $60,000$ on a low-index silica 
platform. The Rowland-type AWG suffers from a nonuniformity of the spectral resolution across the flat image plane due to defocus aberration, causing the spectral resolving power to decrease substantially towards the centre of the 
spectral image. Defocus-corrected AWG-designs using a field-flattening lens in the free propagation region (FPR) have been proposed as a solution \cite{Akca:12}, which, however, adds complexity to the device fabrication. 
An elegant alternative is provided through the three-stigmatic-point geometry, which achieves field-flattening by modifying the grating shape and array waveguide lengths \cite{Wang:01, Lu:03, Shi:05}. Working field-flattened devices 
using three stigmatic points have been demonstrated on polymer and SOI platforms \cite{Lu:05, Bai:10}. This work is the second paper in a series covering the design of integrated diffractive elements for the "Potsdam Arrayed 
Waveguide Spectrograph" (PAWS) \cite{hernandez2020}, a semi-integrated photonic spectrograph using various AWG designs as well as photonic echelle gratings (PEGs) with a modified focal plane using two stigmatic 
points. 

We present the implementation of the world's first field-flattened AWG devices on a low-refractive-index-contrast silica platform 
($\Delta = 0.02$) designed specifically for mid-to high-resolution spectroscopy in astronomy. The new anastigmatic AWG designs are a modification of the canonical Rowland-type Generation-I designs with spectral resolving powers up to 
$R=\lambda/\Delta\lambda=34,800$ at $1550$ nm. In paper I, higher resolving powers were found impractical to achieve experimentally without phase error trimming and were subsequently avoided in the following generation 
of devices. Stigmatic versions of the Gen-I designs were obtained by defining three stigmatic points on a straight line perpendicular to the principal optical axis, located a focal distance $L_f$ above the grating centre. The general properties, 
such as waveguide core width $w=3.4\,\upmu$m,  grating pitch $D=10.57\,\upmu$m and free spectral range (FSR) were retained in order to minimize the required geometry changes. Like their Rowland-counterparts, the anastigmatic AWGs were designed 
to operate in a region of the astronomical H-band between $1500$ nm and $1700$ nm. Fabrication of the designs was accomplished in collaboration with the photonics foundry Enablence USA Components Inc. The devices were fabricated using UV-photolithography and 
atmospheric pressure chemical vapour deposition (APCVD) of $SiO_2$ on a substrate (silica-on-silicon). In this work, we present the construction and simulation of anastigmatic AWG designs operating at a central wavelength of $1550$ nm, as well as 
experimental characterization results of fabricated anastigmatic AWG designs. We show that defocus of the spectral image occurs at off-centre wavelengths due to a difference in dispersion between the waveguide array and free propagation 
region of the anastigmatic AWG. We provide a theoretical condition for defocus cancellation in the wavelength range of operation.
 
\section{Theory}
Aberration theory of the arrayed waveguide grating \cite{Wang:01} provides a simple and versatile method for designing various types of AWGs and PEGs with suppressed optical aberrations.
Designs known as field-flattened AWGs and two-stigmatic-point PEGs \cite{Horst:09} are of special interest for integrated spectroscopy applications due to their uniformity of the point spread function (PSF) across the entire image plane. Low aberration 
geometries can be derived by defining a set of at most three stigmatic points (i.e. points of perfect constructive interference) and their respective wavelengths in the image plane. 
\begin{figure}[!ht]
	\centering
	\includegraphics[width=110mm]{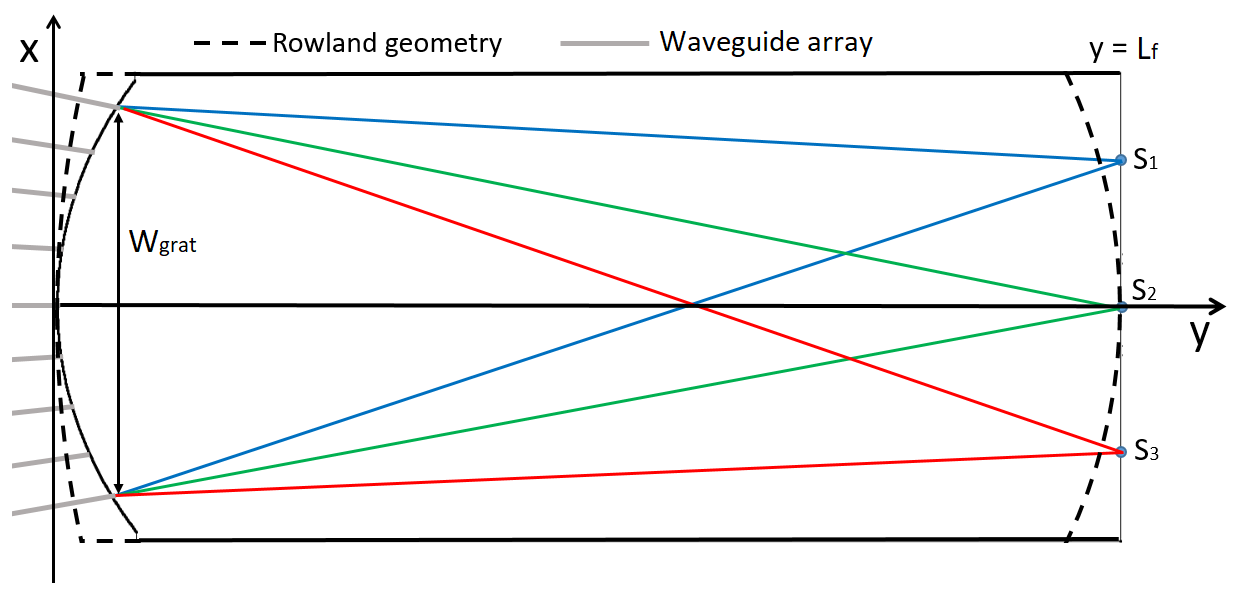}
	\caption{ Schematic layout of a field-flattened star coupler with three stigmatic points $S_1$, $S_2$ and $S_3$ corresponding to shorter and longer wavelengths. Light emerges from the waveguide array on the left side; optimal focus of the spectral image at $y=L_f$.}
	\label{fig:stigmatic_fpr}
\end{figure}
Figure \ref{fig:stigmatic_fpr} schematically illustrates the structure of a field-flattened AWG star coupler. The stigmatic points are arranged in a straight line, minimizing defocus in the entire region between $S_1$ and $S_3$. Note the
non-circular shape of the array-FPR interface on the left side.

A stigmatic-point device geometry is derived by forcing all aberration coefficients in the stigmatic points to zero.
The optical path length function (OPF) and its differential coefficients are given by the set of equations \cite{Wang:01}
\begin{subequations}\label{eq:OPF_equations}
	\begin{equation}\label{eq:opf}
		F(x)=N_s(\lambda) \left[ r_A(x)+r_B(x)\right]+N_a(\lambda)L(x)+m\lambda G(x),\\
	\end{equation}
	\begin{equation}\label{eq:opf_diff}
		0=F^{(n)}(0)=N_s(\lambda)\left[ r_A^{(n)}(0)+r_B^{(n)}(0)\right]+N_a(\lambda)L^{(n)}(0)+m\lambda G^{(n)}(0).\\
	\end{equation}
	\begin{equation}\label{eq:r_diff0}
		r_i^{(0)}(0)=(x_i^2+y_i^2)^{\frac{1}{2}},\\
	\end{equation}
	\begin{equation}\label{eq:r_diff1}
		r_i^{(1)}(0)=-x_i/r_i^{(0)}(0),\\
	\end{equation}
	\begin{equation}\label{eq:r_diff2}
		r_i^{(n)}(0)=\Psi_i-\frac{y_i}{r_i}u^{(n)}(0),\\
	\end{equation}
	\begin{equation}\label{eq:psi}
		\Psi_i=\frac{1}{2r_i^{(0)}(0)}\left[\sum_{k=1}^{n-1}\binom{n}{k}\left(u^{(k)}(0)u^{(n-k)}(0)-r_i^{(k)}(0)r_i^{(n-k)}(0)+2\delta(n-2)\right)\right],\\
	\end{equation}
\end{subequations}
where $x_i$ and $y_i$ are coordinates of the $i^{\text{th}}$ stigmatic point, $N_s$ and $N_w$ are the slab waveguide and array waveguide effective index, respectively, $r_A$ and $r_B$ are geometrical lengths of the paths in the 
input/output FPR (denoted A and B) connecting the stigmatic points with the waveguide array, $L(x)$ is the array waveguide length function, $G(x)$ is the grating period chirp function, $u^{(n)}(0)$ and $r_i^{(n)}(0)$ are the differential coefficients of 
the grating curve function $u(x)$ and the free propagation lengths $r_i(x)$ at $x=0$, respectively, and $m$ is the grating order of the waveguide array. The optical path function $F(x)$ is required to be constant in the range 
$-W_{\text{grat}}/2 < x < W_{\text{grat}}/2$, where $W_{\text{grat}}$ is the width of the grating (see Figure \ref{fig:stigmatic_fpr}), which implies that its derivatives with respect to $x$ must be zero at $x=0$. We introduce 
an additional degree of freedom into the aberration theory by Wang et al. by allowing distribution of the three stigmatic points over different spectral orders. Introducing the reduced effective index $N_i=N_s(\lambda_i)/N_a(\lambda_i)$ 
and the reduced stigmatic wavelength $\Lambda_i=m_i\lambda_i/N_a(\lambda_i)$, equation \eqref{eq:opf_diff} can be written as
\begin{equation}\label{eq:opf_diff2}
	N_i\left[ r_A^{(n)}(0)+r_i^{(n)}(0)\right]+L^{(n)}(0)+\Lambda_i G^{(n)}(0)=0.
\end{equation}
The geometry of the optical paths in the AWG is uniquely defined by three stigmatic points, which are required to exactly satisfy \eqref{eq:opf_diff2}.

Each stigmatic point is defined by its coordinates $(x_i, y_i)$, the corresponding wavelength $\lambda_i$ and the spectral order $m_i$ of the stigmatic point. Inserting the stigmatic point properties into \eqref{eq:opf_diff2}, together with \eqref{eq:r_diff2}, yields the differential coefficients of the 
grating curve function $u(x)$
\begin{equation}\label{eq:u_diff}
	u^{(n)}(0)=\frac{N_1(1-\Lambda)(\Psi_A+\Psi_1)+\Lambda N_3 (\Psi_A+\Psi_3)-N_2(\Psi_A+\Psi_2)}{N_1 (1-\Lambda)\left(\frac{x_A}{r_A}+\frac{x_1}{r_1}\right)+\Lambda N_3\left(\frac{x_A}{r_A}+\frac{x_3}{r_3}\right)-N_2\left(\frac{x_A}{r_A}+\frac{x_2}{r_2}\right)},
\end{equation}
with $\Lambda = (\Lambda_1-\Lambda_2)/(\Lambda_1-\Lambda_3)$. Once the coefficients $u^{(n)}(0)$ are known, the coefficients of $L(x)$ and $G(x)$ follow immediately from equation \eqref{eq:opf_diff2}. 
The differential coefficients are recursively computed from equations \eqref{eq:psi}-\eqref{eq:u_diff} using the condition $u^{(0)}(0)=u^{(1)}(0)=0$ and equations \eqref{eq:r_diff0}-\eqref{eq:r_diff2}. 
The classical Rowland-geometry emerges as a special case when the stigmatic points are located on a circle. The minor generalization of variable spectral orders introduced in this work enables an alternative 
way of defining the AWG geometry by specifying three stigmatic points in three different spectral orders at one single wavelength as opposed to a definition by three different wavelengths in one single order. 
In this work, however, the latter approach was implemented.
			
\section{Design of field-flattened AWG devices}
We begin the design process by defining the target material platform to be a low-refractive-index (RI) 
silica-on-silicon platform with an RI contrast $\Delta = (n_{\text{core}}^2 - n_{\text{clad}}^2)/2 n_{\text{core}}^2 = 0.02$,
cladding refractive index $n_{\text{clad}}=1.444$ and core refractive index $n_{\text{core}}=1.4738$ at $\lambda=1550 $ nm.
The waveguide structure used in this work is identical to the structure described in paper I, with a guiding layer
thickness of $3.4\,\upmu$m and a waveguide core width of $3.4\,\upmu$m - $9.07\,\upmu$m. Waveguide mode fields and wavelength-dependent 
effective indices of the waveguide core were determined by numerical simulations using a commercial implementation (RSoft) of a full-vectorial 
3D-beam propagation method (BPM). The simulated optical properties of the basic building blocks, such as channel waveguides, bends and tapers,
 have been covered in detail in paper I. In summary, we have defined the fundamental design parameters 
as listed in Table \ref{tab:system_specs}.
\begin{table}
	\caption{\label{tab:system_specs} Fundamental structural parameters of the AWG designs.}
		\centering
		\begin{tabular}{c| c c c c c c c} 
		 	\hline
			\textbf{ Material system }& Ge-doped $SiO_2$, $\Delta=0.02$\\ [0.5ex] 
			 \textbf{Core thickness} & $3.4\,\upmu$m \\ 
			 \textbf{Core width (single mode)} & $3.4\,\upmu$m\\
			 \textbf{Core width (taper)} & $9.07\,\upmu$m\\
			 \textbf{Bend radius} & $\geq 1.5$ mm \\
			 \textbf{Focal length} & $5$ mm - $13.5$ mm \\ [1ex] 
			 \textbf{Grating pitch} & $10.57\,\upmu$m \\			
			 \hline
		\end{tabular}
	\end{table}

\subsection{Definition of stigmatic points}
In paper I, we have studied canonical Rowland-type AWG designs with target resolving powers of up to $60,000$. This work focuses on AWG designs with modest resolving powers of up to 35,000 to avoid excessive degradation 
of the PSF due to phase errors. In this work, we have studied eight different custom-designed and fabricated AWGs with FSRs of $11.9$ nm - $47.2$ nm, covering a 
range of spectral resolving powers of $R=11,500$ - $34,800$. Since the spot size of the input waveguide and the angular dispersion of the waveguide grating are fixed by the design, the spectral resolution is determined only by
the FSR and focal length $L_f$.  

Stigmatic points define the locations and associated wavelengths of zero aberration in the focal plane of the AWG and are represented by $S_1 = (-W/2, L_f)$ at $\lambda=\lambda_0 - FSR/2$, $S_2 = (0, L_f)$ at $\lambda=\lambda_0$, $S_3 = (W/2, L_f)$ 
at $\lambda=\lambda_0 + FSR/2$, where $W$ is the width of the dispersed spectral image in the focal plane and $\lambda_0$ is the central wavelength. The three stigmatic points define a focal plane perpendicular to the AWG's optical axis at a distance $L_f$ 
from the centre of the grating. We have determined $W$ for each design by iterative tuning of the parameter and computing of Equations \eqref{eq:OPF_equations} until the grating pitch condition $G^{(1)}(0)=1/D$ was satisfied. 

We have generated AWG designs with focal lengths of $5 $ mm, $7 $ mm, $9 $ mm, $12.728 $ mm and $13.5 $ mm, enumerated 1 - 5, and FSRs $11.9$ nm, $16.1$ nm, $22.6$ nm, $32.2$ nm and $47.2$ nm, enumerated A - E. 
In this work, we have implemented the FSR-$L_f$ combinations A-1, A-2, B-3, B-5, C-4, D-3 and E-5. Each AWG was designed for a central wavelength $\lambda_0=1550$ nm. The design parameters obtained for these eight combinations are 
listed in Table \ref{tab:stigmatic_para}.

\begin{table}
	\caption{\label{tab:stigmatic_para} Design parameters for the definition of stigmatic points $S_1$ - $S_3$.}
		\centering
		\begin{tabular}{c c c c c c c c c} 
		 	\hline
			 Design & A-1 & A-2 & B-3 & B-5 & C-4 & D-3 & E-5 \\ [0.5ex] 
			\hline
			$FSR$ (nm) & 11.9 & 11.9 & 16.1 & 16.1 & 22.6 & 32.2 & 47.2 \\ 
			$L_f$ (mm) & 5 & 7 & 9 & 13.5 & 12.728 & 9 & 13.5 \\ 
			$W$ ($\upmu$m) & 493.2 & 690.5 & 888.0 & 1332.0 & 1254.9 & 893.4 & 1332.0 \\ 
			\hline
		\end{tabular}
	\end{table}

\subsection{Construction of the field-flattened AWGs}
Equations \eqref{eq:OPF_equations}-\eqref{eq:u_diff} were solved numerically for each configuration of stigmatic points, yielding the differential coefficients of the Taylor series expansions of the grating curve $u(x)$, array waveguide length $L(x)$ and grating chirp function $G(x)$.
Grating point coordinates, i.e. locations of waveguides at the array-FPR boundary, and corresponding array waveguide lengths were obtained by evaluating the Taylor polynomials of $u(x)$ and $L(x)$ on a discrete set of points $\left\{ x | x\in \mathbf{R}, G(x)\in \mathbf{Z} \right\}$ in the range $-W_{\text{grat}}/2<x<W_{\text{grat}}/2$.
The AWG devices in this work have been designed with grating aperture angles of $25.6^{\circ}$ - $27.5^{\circ}$, resulting in $W_{\text{grat}}=2.191$ mm ($L_f=5$ mm) - $W_{\text{grat}}=5.794$ mm ($L_f = 13.5$ mm).

\begin{figure}[!ht]
	\centering
	\includegraphics[width=130mm]{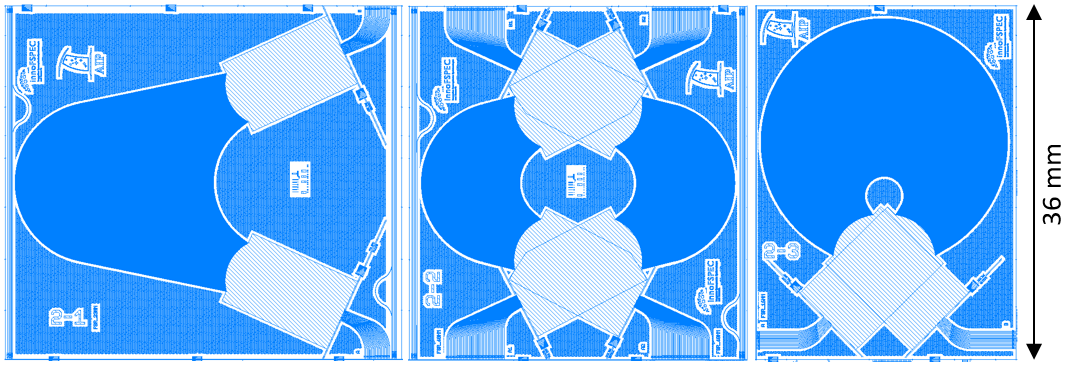}
	\caption{Lithographic mask layouts of AWG designs C-4 (left), E-5 (middle) and B-5 (right).}
	\label{fig:AWG_Designs_Lith}
\end{figure}

The optical delays of the AWGs were realized as conventional three-segmented circular waveguide arrays consisting of two straight segments and one circular segment.
Narrow-FSR AWGs were implemented in a folded configuration of two overlapping FPRs crossing at an angle $\geq 90^{\circ}$. This configuration provides
sufficient room for a consistent waveguide array structure at arbitrarily large path length increments. Wide-FSR AWGs, since they require less space for path length accumulation, were implemented in the traditional horseshoe shape known from commercial telecom devices. 
Arrays of input/output waveguides with a wavelength spacing of $1$ nm were added to each device. The I/O waveguides were not tapered in order to minimize the mode field diameter to achieve the maximum possible spectral resolution.
Lithographic layouts of three selected AWG designs C-4, E-5 and B-5, which represent the main device families, are shown in Figure \ref{fig:AWG_Designs_Lith}. 
The remaining designs are structurally similar. Design D-3 is a modification of E-5 and designs A-1, A-2, B-2 and B-3 are modifications of B-5. The number of array waveguides ranges from 220 in the smallest design A-1 to 579 in designs B-5 and E-5. The AWG designs have foot-prints of $3.45\,\text{cm}^2$ (A-1) - $14.4\,\text{cm}^2$ (C-4).
Layouts of the remaining designs are provided in the supplemental document.

\section{Numerical simulation}\label{sec:num_sim}
The AWG designs were numerically studied for their transmission characteristics using a combination of 3D-BPM and a model of the AWG
utilizing scalar Fraunhofer diffraction in the FPR slab waveguide, which we assume to be homogeneous and isotropic in the plane of propagation.
In the first stage of the simulation, we have used 3D-BPM to obtain the mode field distributions of the untapered input/output waveguides, the tapered array waveguides 
at the FPR-interface and the fundamental propagating mode of the FPR slab. The wavelength-dependent effective indices of the channel waveguides and the slab waveguide
were calculated in the wavelength range 1500 nm - 1700 nm. Material dispersion was included in the model by using refractive index data for $SiO_2$ from the RSoft dielectric 
material library. The calculated mode profiles of the I/O waveguides and tapered array waveguides were transformed into their corresponding far fields in the FPR slab
and subsequently used in the scalar diffraction model. 
\begin{figure}[!ht]
	\centering
	\includegraphics[width=130mm]{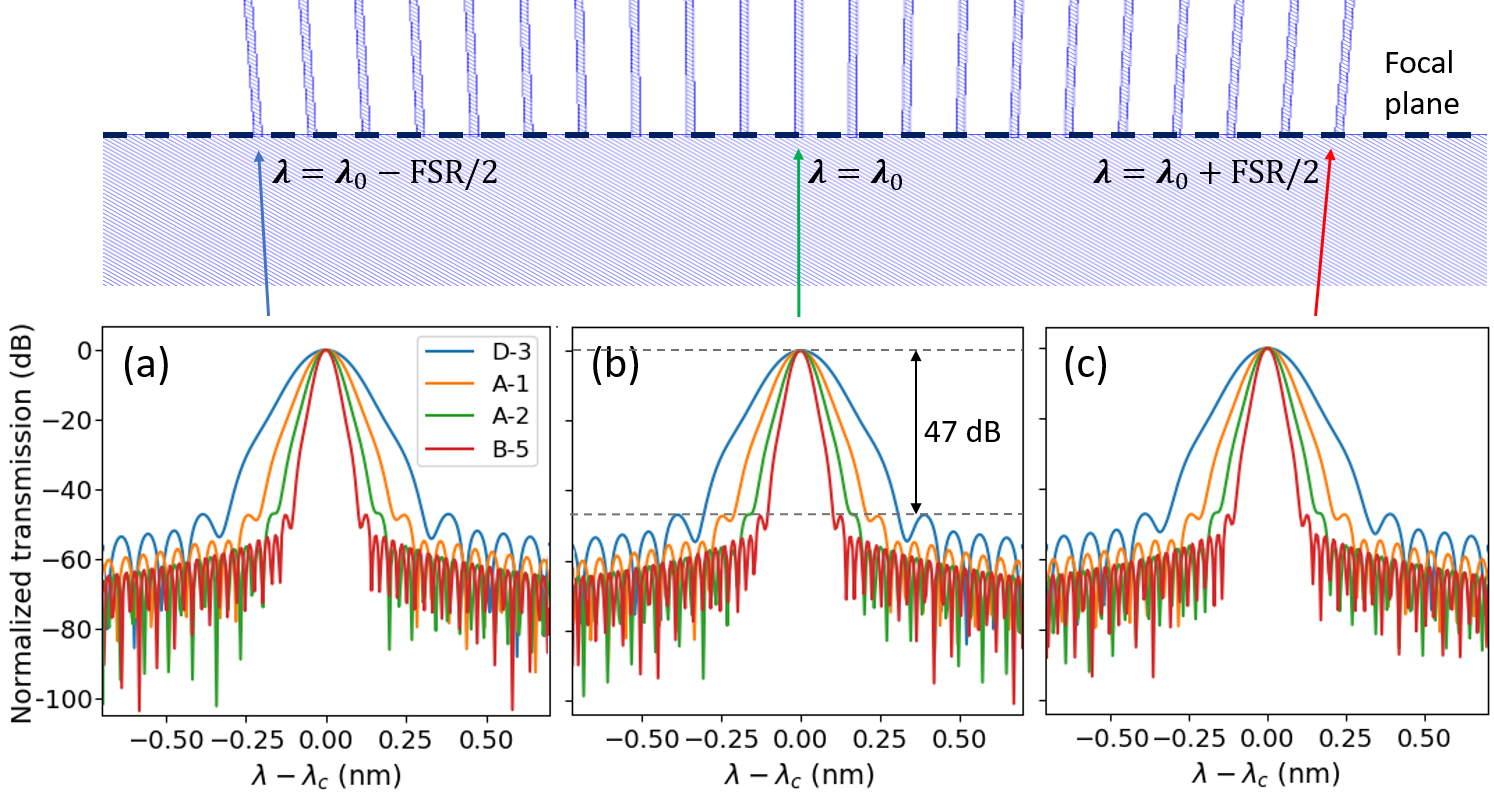}
	\caption{Simulated power transmission curves of designs A-1, A-2, D-3 and B-5. (a) Output channel transmission at the short-wavelength edge ($\lambda_c=\lambda_0-FSR/2$). (b) Transmission of central output channels ($\lambda_c=\lambda_0$). (c) Transmission at the long-wavelength edge ($\lambda_c=\lambda_0+FSR/2$). Wavelength scale relative to central wavelength $\lambda_c$ of the transmission peaks. Top: Array of output waveguides; transmission (a), (b) and (c) simulated for leftmost, central and rightmost waveguides.}
	\label{fig:Sim_transmission}
\end{figure}

The power transmission from input waveguide $j$ to output waveguide $k$ was simulated by numerically computing \cite{Okamoto:2006}
\begin{subequations}
	\begin{equation}\label{eq:AWG_analytical}
		T_{jk}(\lambda)=\left|\sum_{l=1}^{N} c_{jk}(l)\exp\left(-i\Omega_{jk}(l)\right)\right|^2
	\end{equation}
	\begin{equation}
		c_{jk}(l) = f(\sigma_{jl})g(\rho_{jl})g(\rho_{kl})h(\sigma_{kl})
	\end{equation}
	\begin{equation}
		\Omega_{jk}(l) = \frac{2\pi}{\lambda}\left( n_s(r_{jl}+r_{kl})+ n_a L_l \right),
	\end{equation}
\end{subequations}
where $\Omega_{jk}$ is the signal phase after propagation, $f$, $g$ and $h$ are angular far field functions of the input wave\-guides, array wave\-guides and output wave\-guides, respectively; $\sigma_{jl}$ and $\rho_{jl}$ 
are emission angles relative to the axis of the input waveguide $j$ and array waveguide $l$, respectively; $n_a$ and $n_s$ are effective indices of the waveguide array and the FPR slab, respectively; $r_{jl}$ and $r_{kl}$ are 
the distances between the waveguide array and input/output waveguides; and $L_l$ is the length of the $l^{\text{th}}$ array waveguide. 

Transmission spectra were calculated for each AWG design between $1500$ nm and $1680$ nm. The 3-dB bandwidths $\Delta\lambda$ of the transmission curves were 
of special interest, as they provide a rough lower estimate of the spectral resolution. Figure \ref{fig:Sim_transmission} shows simulation results for designs A-1, A-2, D-3 and B-5. Figures \ref{fig:Sim_transmission}(a), (b) and (c) show
the transmission curves of output channels on the short-wavelength edge, the centre and the long-wavelength edge of the output region, respectively. The transmission curves are remarkably similar for edge channels and central channels, 
to the point of being almost indistinguishable, which indicates a very high uniformity of spectral resolution across the flat focal plane. The sidelobe intensity is -47 dB in all designs.

\begin{figure}[!ht]
	\centering
	\includegraphics[width=130mm]{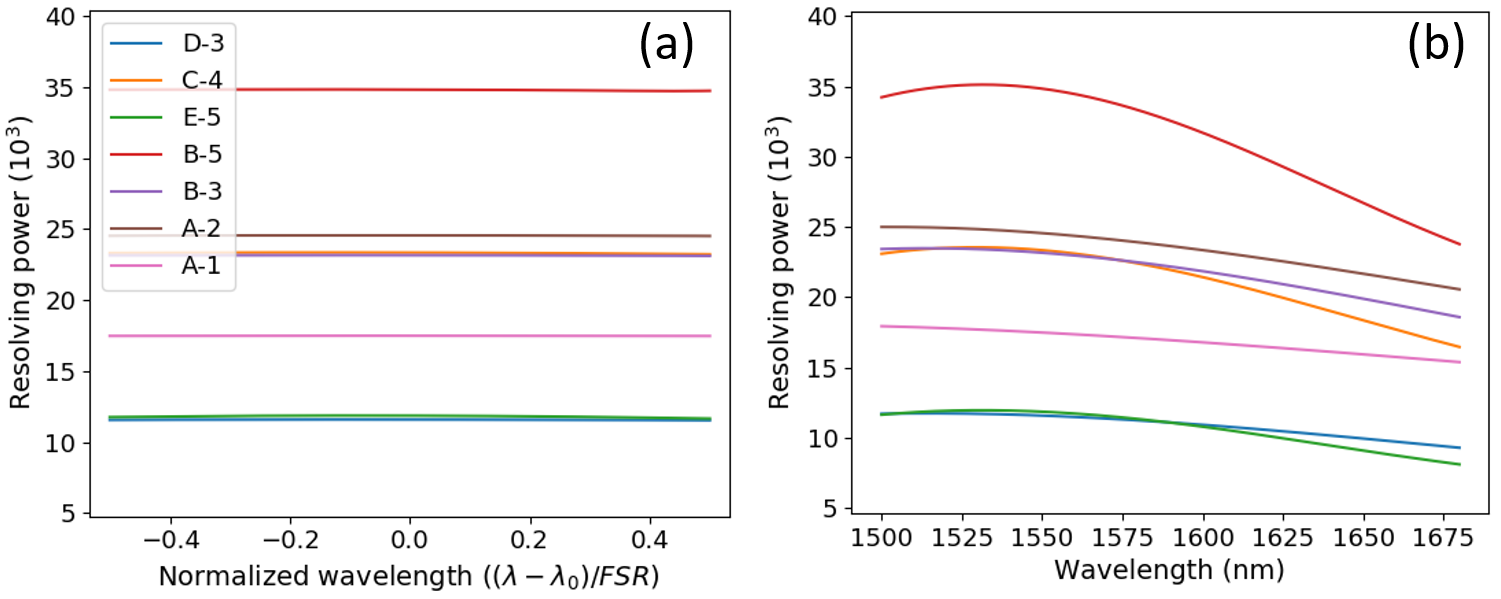}
	\caption{Simulated spectral resolving power of the field-flattened AWG designs. (a) Resolving power in the main spectral order around 1550 nm, spanning one FSR. (b) Variation of resolving power in the entire operating wavelength range.}
	\label{fig:Sim_RP}
\end{figure}

We have evaluated the spatial uniformity of the transmission peak bandwidth in the spectral image region of the AWG by simulating the transmission for each of the output waveguides distributed equidistantly across the focal plane and measuring its
3-dB bandwidth in the wavelength range $1500$ nm - $1680$ nm. Uniformity of the transmission bandwidths of each channel across the output waveguide array indicates uniformity of the focused beam profile in the focal plane. Figure \ref{fig:Sim_RP}(a) shows the spectral resolving power measured across all output waveguides in the wavelength interval $(\lambda_0-FSR/2, \lambda_0+FSR/2)$, whereby each waveguide transmits a particular wavelength depending on its location in the focal plane (see top of Figure \ref{fig:Sim_transmission}). The resolving power is nearly constant across one FSR in all designs. In the main spectral order, simulated variations of $R=\lambda/\Delta\lambda$, defined as $\Delta R = (R_{\text{max}}-R_{\text{min}})/R_{\text{max}}$ 
were found to be on the order of $10^{-4}$ - $10^{-2}$. Designs with shorter focal lengths exhibit less variation of $R$ in comparison with larger AWGs with longer focal lengths. 
The stigmatic point method ensures a low variation of spectral resolving power in the main spectral order spanning one FSR. In practice, however, the AWG devices are expected to operate in a wide wavelength range of a few $100 $ nm, covering multiple spectral orders. 
Figure \ref{fig:Sim_RP}(b) shows the variation of spectral resolving power between $1500$ nm and $1680$ nm. A degradation of spectral resolution is observed at off-centre wavelengths. Again, the severity of $R$ non-uniformity increases with focal length, being strongest in design 
B-5 with $L_f = 13.5$ mm.

Table \ref{tab:sim_results} presents the simulation results for all seven AWG designs. We have observed simulated 3-dB transmission bandwidths of $44.5$ pm - $133.9$ pm, corresponding to resolving powers of 11,500 - 34,800 at 1550 nm.
\begin{table}[!ht]
	\caption{\label{tab:sim_results} Simulated 3-dB channel transmission bandwidths and estimated spectral resolving powers at 1550 nm.}
	\centering
	\begin{tabular}{c c c c c c c c c} 
	 	\hline
		 Design & A-1 & A-2 & B-3 & B-5 & C-4 & D-3 & E-5 \\ [0.5ex] 
		\hline
		$\Delta\lambda$ (pm) & 88.7 & 63 & 66.9 & 44.5 & 66.4 & 133.9 & 130.8 \\ 
		est. $R$ & 17,400 & 24,600 & 23,100 & 34,800 & 23,300 & 11,500 & 11,800 \\ 
		\hline
	\end{tabular}
\end{table}

\subsection{Chromatic defocus in field-flattened AWGs}
The analysis of the simulated transmission presented in the previous section shows a significant increase of transmission peak bandwidth, respectively spectral resolving power $R$ at off-centre wavelengths, which becomes more pronounced in AWG designs with larger star couplers. 
			This type of behaviour is very critical for high-resolution astronomical applications due to the requirement of operation in a wide wavelength range. Strong non-uniformity of the diffraction peak width across the spectral orders severely limits the
			usable wavelength range at high resolutions. In this section, it is shown that the observed degradation of the PSF in anastigmatic AWG designs is caused by the difference of waveguide array and star coupler slab waveguide dispersion and a condition 
			for zero-chromatic-defocus is obtained.

\begin{figure}[!ht]
	\centering
	\includegraphics[width=80mm]{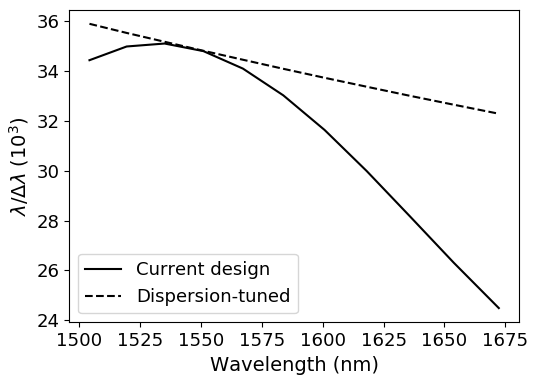}
	\caption{Simulated spectral resolving power of design B-5. Comparison between unmodified dispersion (solid) and a hypothetical dispersion-tuned design (dashed).}
	\label{fig:Chroma_Defocus}
\end{figure}

			The condition for constructive interference at the central point on the focal plane is defined as
			\begin{equation}\label{eq:condition_interference}
				n_a l_k + 2n_s s_k = L_0 + km\lambda_0,	
			\end{equation}
			where $\lambda_0$ and $m$ are the central wavelength and grating order of the AWG, respectively; $n_a$ and $n_s$ are the array waveguide effective index and the slab waveguide effective index, respectively; $l_k$ is the length of the $k^{th}$ array waveguide; 
			$s_k$ is the propagation length between the $k^{th}$ array waveguide and the centre of the focal plane; $L_0=n_a l_0 + 2n_s s_0$ is the total optical length of path $k=0$.
			The waveguide effective index of silica channel waveguides and slab waveguides in the studied wavelength range can be approximated by
			\begin{equation}\label{eq:neff_linearization}
				n_{a,s}(\lambda) \approx n_{a,s}(\lambda_0)+(\lambda-\lambda_0)\frac{dn_{a,s}}{d\lambda}(\lambda_0).
			\end{equation}
			In equation \eqref{eq:condition_interference}, the geometrical lengths of the array and slab sections in the anastigmatic case can be expressed as $l_k = l_0 + k\Delta l + \delta l_k$ and $s_k = s_0 + \delta s_k$, where $l_0 + k\Delta l$ and $s_0$ represent the array 
			waveguide length and free propagation length in the Rowland-geometry, respectively, while $\delta l_k$ and $\delta s_k$ represent the length deviations from the Rowland geometry. 
			With the approximation \eqref{eq:neff_linearization} and $n_a(\lambda_0)\Delta l=m\lambda_0$, the constructive interference condition \eqref{eq:condition_interference} becomes
			\begin{subequations}\label{eq:condition_interference_2}
				\begin{equation}\label{eq:condition_interference_2_1}
					n_a(\lambda_0)\delta l_k + 2n_s(\lambda_0)\delta s_k + (\lambda-\lambda_0)\left[\frac{dn_a}{d\lambda}(k\Delta l + \delta l_k) + 2\frac{dn_s}{d\lambda} \delta s_k \right]=L_0-C,
				\end{equation}
				\begin{equation}
					C= n_a(\lambda_0) l_0 + 2n_s(\lambda_0) s_0 + (\lambda-\lambda_0)\left[ \frac{dn_a}{d\lambda}l_0 + 2\frac{dn_s}{d\lambda} s_0 \right].
				\end{equation}
			\end{subequations}
			Evaluation of equations \eqref{eq:condition_interference_2} at $\lambda=\lambda_0$ yields the relation between the modulations of the geometrical path lengths as $n_a(\lambda_0)\delta l_k + 2n_s(\lambda_0) \delta s_k = 0$.
			Finally, equation \eqref{eq:condition_interference_2_1} becomes
			\begin{equation}\label{eq:interference_condition_final}
				(\lambda-\lambda_0)\left[ \frac{dn_a}{d\lambda} k \Delta l +\left (\frac{dn_a}{d\lambda}-\frac{n_a(\lambda_0)}{n_s(\lambda_0)}\frac{dn_s}{d\lambda}\right)\delta l_k \right ]=L_0-C.
			\end{equation}
			It is immediately clear that the above equation only holds in the cases (i) $\lambda=\lambda_0$ and (ii) $dn_a/d\lambda=dn_s/d\lambda=0$, as the right hand side is constant with respect to $k$.
			At off-centre wavelengths, waveguide dispersion gives rise to aberrations, which are represented by the two terms in square brackets on the left hand side of equation \eqref{eq:interference_condition_final}.
			The first term depends only on the variation of the waveguide array effective index and adds a constant excess path length increment to the array waveguide lengths, which causes a shift of the central wavelength.
			The second term depends on the variation of both the waveguide array and slab waveguide effective index and is proportional to the modulation $\delta l_k$ of the array waveguide lengths. In an anastigmatic AWG design, $\delta l_k$ is 
			non-linear and symmetric with respect to the centre of the grating. The aberrations resulting from the second term at off-centre wavelengths tend to defocus the image.
			The condition for zero chromatic defocus is obtained from equation \eqref{eq:interference_condition_final} as
			\begin{equation}\label{eq:zero_defocus_condition}
				Q=\frac{dn_a}{d\lambda}-\frac{n_a}{n_s}\frac{dn_s}{d\lambda}=0\,{\upmu \text{m}}^{-1}.
			\end{equation}
The effect of condition \eqref{eq:zero_defocus_condition} on the uniformity of spectral resolving power is shown in Figure \ref{fig:Chroma_Defocus}.
For our current designs, numerical simulations indicate $n_s>n_a$ and $dn_a/d\lambda>dn_s/d\lambda$ and $Q=-5.5\times 10^{-3}\,{\upmu \text{m}}^{-1}$.
This analysis shows several ways to mitigate or eliminate chromatic defocus. It follows from Equation \eqref{eq:interference_condition_final} that small values of $\delta l$ result in smaller aberrations. Therefore, designs with smaller foot-print are advantageous in terms of resolving power uniformity. Alternatively, waveguide cores with
a large aspect ratio, such as strip waveguides, can be used to achieve a smaller $Q$ value. A promising candidate for this purpose is the SiN based TriPleX technology \cite{Worhoff:15, Roeloffzen:18}.  Another possibility is the fabrication of AWGs with different refractive indices for the array waveguides and slab waveguides, such that condition \eqref{eq:zero_defocus_condition} is satisfied.

\section{Fabrication and testing}\label{sec:fab_testing}
\begin{figure}[!ht]
	\centering
	\includegraphics[width=130mm]{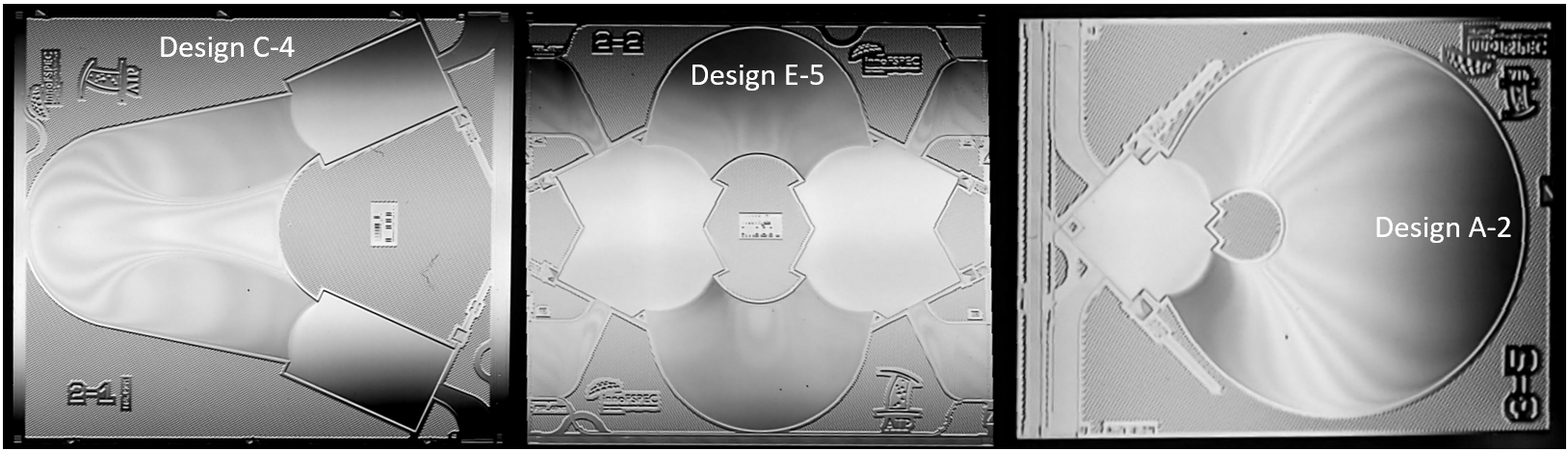}
	\caption{Fabricated AWG chips C-4 (left), E-5 (centre) and A-2 (right).}
	\label{fig:AWG_fab}
\end{figure}

Like their predecessor generation, the field-flattened AWG designs have been fabricated by an external foundry on a $\Delta=2\%$-contrast silica-on-silicon material platform using UV-photolithography and APCVD technology. A $3.4\,\upmu$m thick Ge-doped silica guiding layer was deposited on top of a $20\,\upmu$m thick thermal oxide buffer, followed by application of the contact mask and reactive ion etching (RIE) of the waveguide structure. Finally, a $15\,\upmu$m thick top cladding layer was deposited. 
Figure \ref{fig:AWG_fab} shows photographs of three fabricated chips of type C-4, E-5 and A-2.
The fabricated AWG chips were tested in our laboratory using a T100S-HP CL-band tunable laser source (TLS) and a CT440 component tester unit. The test equipment was interfaced with the AWG chips using butt-coupling of standard SMF-28 fibres and the measurements were normalized with the transmission of a straight reference waveguide. Power transmission spectra were measured between $1500$ nm and $1680$ nm in continuous-sweep mode with a sampling period of $2$ pm and a relative accuracy of $\pm 1$ pm on multiple output waveguides located in the central, peripheral and intermediate regions of the spectral image. 

\begin{figure}[!ht]
	\centering
	\includegraphics[width=130mm]{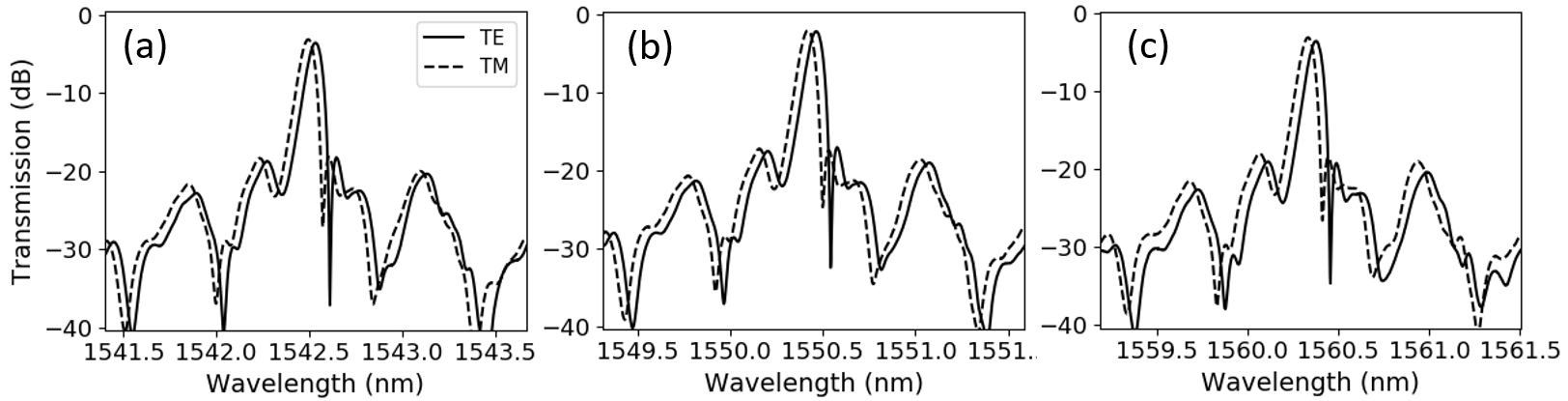}
	\caption{Measured TE and TM transmission of edge channels (a, c) and central channel (b) of AWG C-4.}
	\label{fig:Meas_Trans_2-1}
\end{figure}

The first group of devices under test consists of low-order horseshoe designs C-4, D-3 and E-5. These devices show central channel insertion losses of $2.2$ dB (C-4), $1.66$ dB (D-3) and $1.65$ dB (E-5). We have measured 3-dB bandwidths of $77.5$ pm, $146$ pm and $152$ pm, respectively. The designs achieve resolving powers of $20,000$ (C-4), $10,600$ (D-3) and $10,200$ (E-5) with TE-polarized light, reaching $87\%$, $92\%$ and $86\%$ of their theoretical resolving power, and the transmission curves exhibit polarization dependent wavelength shifts $PD\lambda$ of $40$ pm, $48$ pm and $65$ pm, respectively. The polarization splitting of the transmission spectrum implies a reduced spectral resolution with unpolarized input signals. 
%We have estimated the unpolarized resolving power by numerically adding the measured TE and TM transmission curves and measuring the bandwidth of the combined TE and TM transmission. The resolving power is reduced by $20\%$ in device C-4, $7.5\%$ in device D-3 and $5.9\%$ in device E-5. The severity of resolving power degradation increases with the ratio $PD\lambda/\Delta\lambda$. 
Figure \ref{fig:Meas_Trans_2-1} shows the measured transmission curves of the edge and centre output channels for the highest-resolution device C-4 in the group. Results for devices D-3 and E-5 are shown in the supplemental document. 
\begin{figure}[!ht]
	\centering
	\includegraphics[width=130mm]{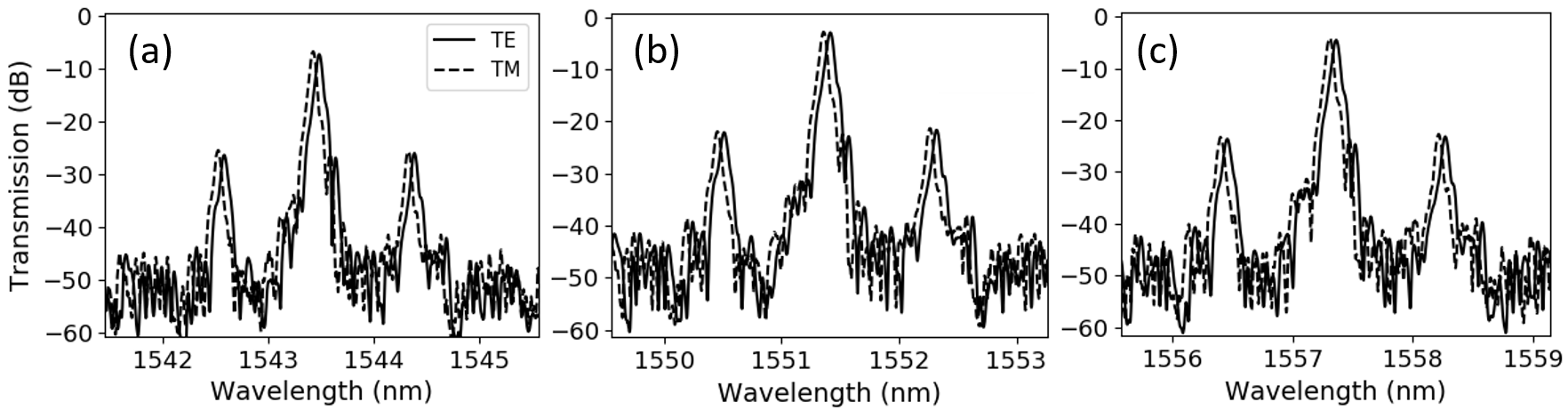}
	\caption{Measured TE and TM transmission of edge channels (a, c) and central channel (b) of AWG B-5.}
	\label{fig:Meas_Trans_2-3}
\end{figure}
\begin{figure}[!ht]
	\centering
	\includegraphics[width=130mm]{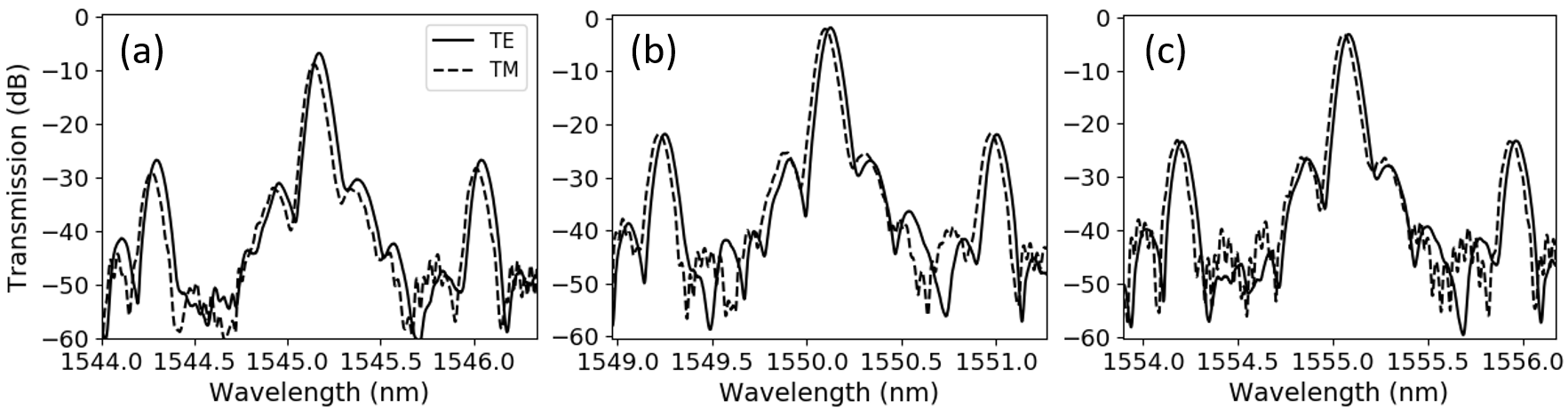}
	\caption{Measured TE and TM transmission of edge channels (a, c) and central channel (b) of AWG A-2.}
	\label{fig:Meas_Trans_3-2}
\end{figure}

The second group of AWGs comprises two high-order devices of the folded type with a FSR of $16.1$ nm (B-3 and B-5). The measured central-channel insertion loss of these devices is $1.99$ dB (B-3) and $3$ dB (B-5). With 3-dB bandwidths of $78$ pm and $56$ pm, respectively, these designs achieve $R=20,000$ (B-3) and $R=27,600$ (B-5) for the TE polarization, reaching $86.6\%$  and $79.3\%$ of their theoretical performance. Polarization dependent wavelength shifts were found to be $39$ pm in device B-3 and $57$ pm in device B-5. Device B-5 is most severely affected by polarization dependence, as $PD\lambda \approx \Delta\lambda$ in this case. 
%Operation with unpolarized light reduces the resolving power by $13.6\%$ in device B-3 and $49.3\%$ in device B-5. 
 Figure \ref{fig:Meas_Trans_2-3} shows the measured transmission curves of AWG B-5. The transmission curves show peculiar features in the form of two symmetrical sidelobes accompanying the main transmission peak with a crosstalk level of $-19$ dB relative to the transmission maximum. The nature of the anomaly indicates two possible causes, which can be (i) a transversal periodic modulation of the waveguide array amplitude profile and (ii) a transversal periodic modulation of the waveguide array phases due to systematic fabrication errors. The exact origin of the sidelobe anomaly in the folded-type design is as yet unknown and will be investigated in a future study.

The third group of AWGs consists of devices A-1 and A-2, which are modified B-x type AWGs with a reduced FSR of $11.9$ nm. The central-channel insertion loss of the fabricated designs is $1.45$ dB in device A-1 and $2$ dB in device A-2. The devices have transmission peak 3-dB bandwidths of $96$ pm and $70$ pm, or, $R=16,100$ and $R=22,000$, respectively, corresponding to $92.5\%$ and $89.4\%$ of the theoretical resolving power. The polarization sensitivity is lowest in these two devices, with $PD\lambda=28$ pm (A-1) and $PD\lambda=27$ pm (A-2). 
%The estimated degradation of resolving power in the unpolarized case is $6.8\%$ in device A-1 and $13.6\%$ in device A-2. The highest estimated resolving power in the unpolarized case is observed in device A-2 with $R=19,000$ at $1550$ nm. 
Figure \ref{fig:Meas_Trans_3-2} shows the measured transmission of AWG A-2. Similarly to the case B-5, this device exhibits symmetrical sidelobes with a crosstalk level of $-20$ dB relative to the transmission maximum. Similar spectral anomalies can be observed in folded AWG designs of the Rowland-type, presented in paper I, which excludes the field-flattening geometry as a possible cause. 

\begin{figure}[!ht]
	\centering
	\includegraphics[width=130mm]{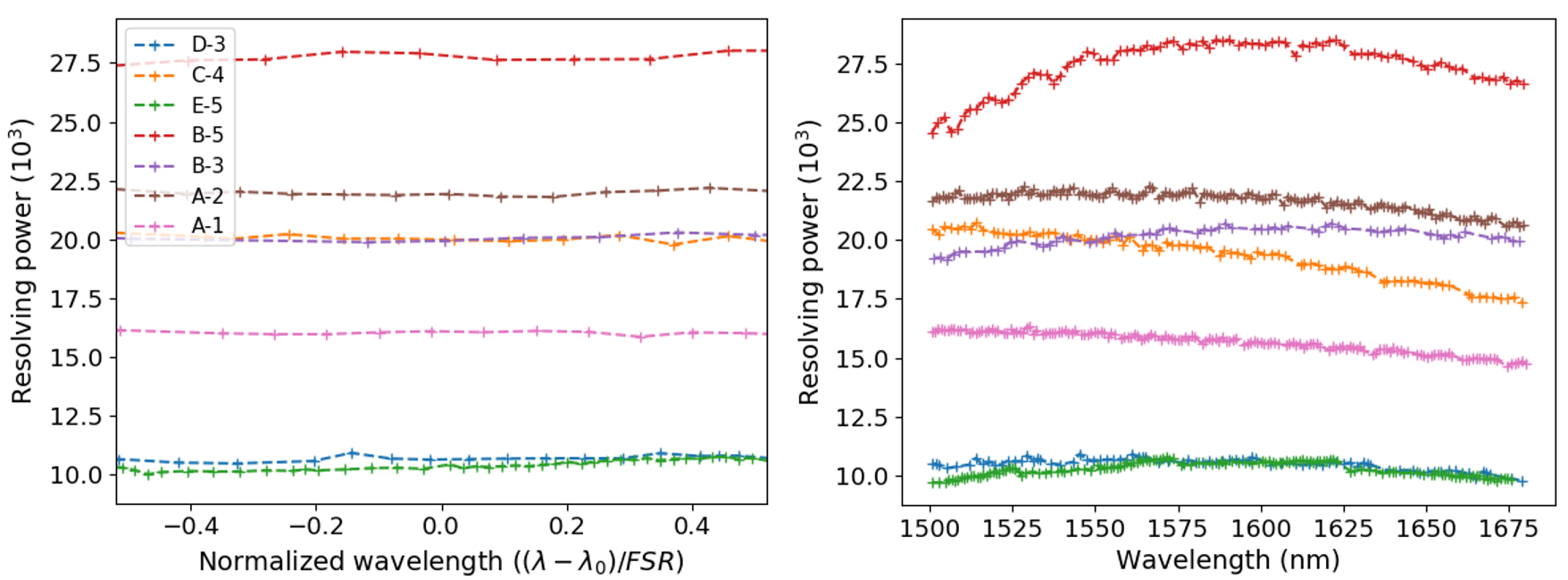}
	\caption{Measured resolving power of the fabricated flat-fielded AWG designs. Variation of resolving power across one FSR (left) and the entire wavelength range of operation (right).}
	\label{fig:RP_Measured}
\end{figure}

In order to test the uniformity of spectral resolving power across the flat image plane of the AWGs, we have measured the power transmission curves of multiple output waveguides distributed equidistantly in the flat focal plane of the output FPR. The array of output waveguides was designed to cover the entire region of the diffraction image, thereby allowing
the characterizaton of the PSF uniformity in the entire focal plane by measuring the transmission curves on each output waveguide in the wavelength range $1500$ nm - $1680$ nm.
After obtaining the measurements for all AWG designs, we have calculated estimates of the spectral resolving power in the entire spectral image region (see Figure \ref{fig:Sim_RP} for comparison), as shown in Figure \ref{fig:RP_Measured} (left). The results show a low variation of the resolving power across the focal plane, with relative variations of $1.7\%$ (A-1), $1.8\%$ (A-2), $2\%$ (B-3), $2.1\%$ (B-5), $2.4\%$ (C-4), $4.1\%$ (D-3) and $5.8\%$ (E-5). The long-range variation of $R$ is shown on the right side of Figure \ref{fig:RP_Measured}. Devices A-2 and B-3 exhibit the smallest long-range variation of $R$ with a total difference of $6\%$ between maximum and minimum in both AWGs, closely followed by devices D-3, A-1 and E-5 with $R$ variations of $7.5\%$, $8\%$ and $8.4\%$, respectively. The strongest variation of $R$ was observed in devices B-5 and C-4 with $14\%$ and $15.6\%$, respectively. The transmission peak bandwidths were measured with an accuracy of $\pm 2$ pm. Correspondingly, the resolving power was measured with an accuracy $\leq 1000$, which is reflected by the measured short-range variation of $R$.

The AWG devices in this work have been characterized by measuring the transmission from an input waveguide to an array of output waveguides, which only shows the properties of the AWGs as multiplexing/demultiplexing optical filters, but does not reveal the focused diffraction image directly. For the purpose of direct imaging of the spectrum by a camera, the output waveguide array will be removed by dicing of the output facet and the output facet will be polished to optical quality in a pending post-processing step. The procedure was performed on two selected AWGs of the previous generation and the results were presented in paper I. In order to predict the spectral resolving power in the direct imaging regime after removal of the output waveguides for the devices in this work, we have conducted a numerical analysis, presented in the next section.

\section{Numerical prediction of the spectral resolving power after dicing} 
In order to obtain the spatial profile of the focused beam at the interface between the FPR and the output waveguides in the fabricated AWGs, we have used the model outlined in section \ref{sec:num_sim} to numerically recreate the measured transmission of the devices. In a Monte-Carlo process, which minimized the residual sum of squares between the output of the model and the measured transmission curves by random variation of the optical path lengths \cite{Stoll1:21, Oh:12}, we have fitted the AWG model with the experimental results, obtaining numerical models which closely match the fabricated AWGs in transmission behaviour.
\begin{figure}[!ht]
	\centering
	\includegraphics[width=130mm]{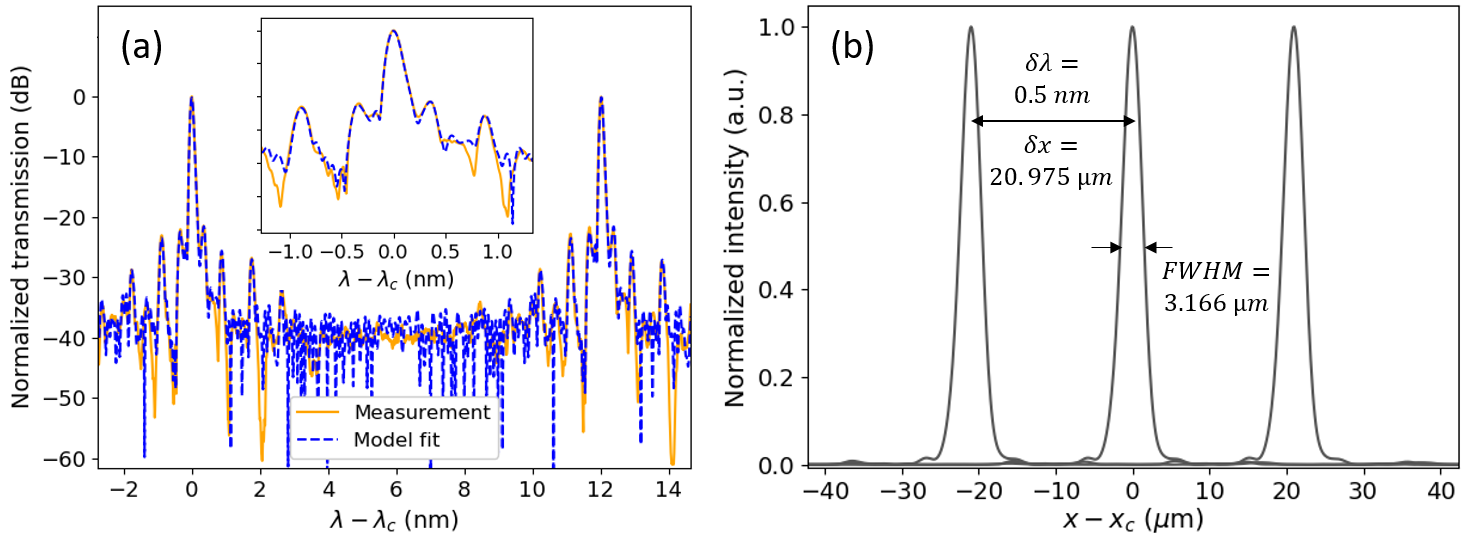}
	\caption{(a) Measured power transmission of the central output waveguide of device A-1, showing two spectral orders (orange), and simulated output of AWG model fitted with measured data (blue dashed). (b) Simulated diffraction image intensity in the focal plane at three wavelengths 0.5 nm apart, centred around 1550 nm. }
	\label{fig:Fit_sim}
\end{figure}
Next, we have used the fitted models to calculate the spatial intensity distribution in the focal plane at several wavelengths for each of the AWG devices. The  wavelength was varied in 0.5 nm steps around the central wavelength 1550 nm. The fitting process was repeated multiple times to ensure stability and the obtained resolving power values were averaged over the multiple runs. A variation of the resolving power estimate in a range $1\%$ - $4\%$ between fitting runs was observed. Figure \ref{fig:Fit_sim} shows the results of model fitting and numerical simulation of the diffraction image of AWG A-1. The wavelength axis is shown relative to the output waveguide transmission maximum $\lambda_c$ and the fitted data region contains transmission peaks of two consecutive spectral orders, ensuring a correct FSR in the fitted model. The measured power transmission on the central output waveguide of the device is shown along with the fitted model output in Figure \ref{fig:Fit_sim}(a). The measured data matches the fit well down to a level of -33 dB, below which discrepancies appear. The quality of the fit is sufficient for an estimate of the full-width-half-maximum (FWHM) of the output beam in the focal plane, as the region of interest of the measured curve is approximated with an accuracy $<0.1$ dB, which is in the range of the experimental measurement accuracy. Figure \ref{fig:Fit_sim}(b) shows the simulated diffraction images in the focal plane of the AWG at incremental wavelengths from 1549.5 nm to 1550.5 nm in 0.5 nm steps. We have obtained estimates of the spectral resolving power in the direct-imaging regime as
\begin{equation}\label{eq:R_image}
	R_{\text{direct}} = \frac{\lambda\cdot\delta x}{\delta\lambda \cdot FWHM},	
\end{equation}
where $\delta\lambda$ is the wavelength shift and $\delta x$ is the corresponding spatial shift of the diffraction image. The estimated resolving power of device A-1 was calculated as 20,540 at 1550 nm, which exceeds the value estimated from the channel transmission measurement by a factor of 1.28. The analysis was performed for all remaining AWGs, each case yielding a higher resolving power than previously estimated from measurement. The lowest predicted $R_{\text{direct}}$ was calculated for device E-5 as 12,880, while the highest values were achieved by devices A-2 and B-5 with 29,490 and 36,000, respectively.

\begin{figure}[!ht]
	\centering
	\includegraphics[width=130mm]{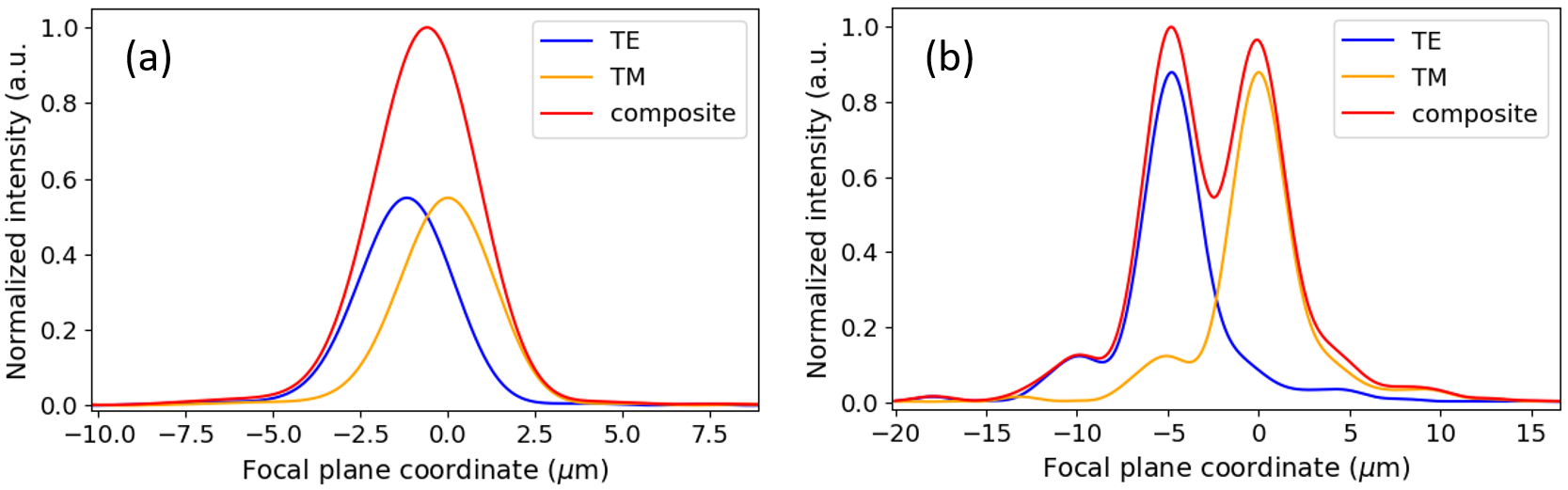}
	\caption{Simulated polarization-shifted PSFs and their composite intensity profile in the focal plane for (a) device A-1, (b) device B-5.}
	\label{fig:pol_psf}
\end{figure}

The above results assume strictly polarized input signals. As mentioned earlier, the resolving capability of the AWG devices suffers from birefringence when operating with unpolarized light. We have evaluated the degradation of resolving power for mixed TE/TM input by using the $PD\lambda$ values obtained in section \ref{sec:fab_testing} to calculate the spatial polarization splitting of the PSF. Assuming fully unpolarized light, we added the intensities of the two polarization components and measured the FWHM of the composite peak. In the case of device B-5, we found that operation in the unpolarized regime is not possible, since the TE/TM components appear as two distinct lines due to the large $PD\lambda$ in this AWG. For the remaining devices, we calculated resolving powers between 9,600 (E-5) and 24,000 (A-2) by inserting the composite FWHM into equation \eqref{eq:R_image}. We found that the resolving power decreased by $9.4\%$ - $35.2\%$ in comparison with the polarized regime. Figure \ref{fig:pol_psf} shows the cases of device A-1 (a), which exhibits the lowest ratio $PD\lambda/\Delta\lambda$, and device B-5 (b) with the largest $PD\lambda/\Delta\lambda$.

The summarized characterization results are presented in Table \ref{tab:test_results}. 
\begin{table}
	\caption{\label{tab:test_results} Summarized characteristics of the fabricated AWG devices.}
	\centering
	\begin{tabular}{c c c c c c c c c} 
		 	\hline
			 Design & A-1 & A-2 & B-3 & B-5 & C-4 & D-3 & E-5 \\ [0.5ex] 
			\hline
			$FSR$ (nm) & 11.9 & 11.9 & 16.1 & 16.1 & 22.6 & 32.2 & 47.4 \\ 
			$IL$ (dB) & 1.45 & 2 & 1.99 & 3 & 2.2 & 1.66 & 1.65 \\
			$\Delta \lambda$ (pm) & 96 & 70 & 78 & 56 & 77.5 & 146 & 152 \\
			$PD\lambda$ (pm) & 28 & 27 & 39 & 57 & 40 & 48 & 65 \\
			$R$ simulated & 17,400 & 24,600 & 23,100 & 34,800 & 23,300 & 11,500 & 11,800 \\
			$R$ measured (TE) & 16,100 & 22,000 & 20,000 & 27,600 & 20,000 & 10,600 & 10,200 \\
			%$R$ unpolarized & 15,000 & 19,000 & 16,000 & 14,000 & 16,000 & 9,800 & 9,600 \\
			$R_{\text{direct}}$ (TE) & 20,540 & 29,490 & 25,270 & 36,000 & 25,760 & 13,880 & 12,880 \\
			$R_{\text{direct}}$ (unpolarized) & 18,600 & 24,000 & 18,300 & - & 16,700 & 12,100 & 9,600 \\
			$\Delta R$ & 0.08 & 0.06 & 0.06 & 0.14 & 0.156 & 0.075 & 0.084 \\
			\hline
	\end{tabular}
\end{table}

\section{Conclusion}
In our second paper in a series of works on integrated photonic spectrograph design, we have presented a new generation 
of optimized low-aberration AWG devices for applications in spectroscopy utilizing the three-stigmatic-point construction method.
In the theoretical part of this work, we have presented a modified version of AWG aberration theory by Wang et al., introducing
the possibility of distributing stigmatic points over multiple spectral orders. We have applied aberration theory of arrayed waveguide gratings 
on existing conventional AWG designs on a $2\%$-contrast silica platform to obtain AWG designs with uniform spectral resolving power 
on a flat focal plane in the near infrared spectral region between $1500$ nm and $1680$ nm, overcoming the defocus aberration of 
conventional AWG designs. The AWG designs reported in this work have theoretical resolving powers between $R=11,800$ and $R=34,800$,
determined numerically. 
Simulations of the AWG designs have shown low variation of spectral resolving power within the main spectral order of the 
AWG on the order of $10^{-4}$ - $10^{-2}$, as well as variations of the resolving power across the entire wavelength range of operation 
of $13.9\%$ in the smallest AWG design A-1 to $30\%$ in design B-5. We have theoretically shown that the observed degradation of resolving 
power is caused by the difference in dispersion between the waveguide array and the free propagation region which causes a chromatic defocus
effect and is a property unique to three-stigmatic-point AWG devices. From the analysis, we have obtained a condition for zero chromatic 
defocus and proposed possible solutions to the issue. 

The AWG designs constructed in our work have been fabricated by an external foundry on a 6-inch silicon 
wafer using APCVD technique. The fabricated devices were tested for their spectral transmission characteristics using a T100S-HP tunable laser
and a CT440 component tester in the wavelength range $1500$ nm - $1680$ nm.

The FSRs of the fabricated devices were in very good agreement with the design values, indicating that material and
geometrical properties of the fabricated waveguide structures are close to the requirements. 
All devices exhibited insertion losses relative to a reference waveguide at or below $3$ dB, the lowest being 
observed in the smallest high-order AWG A-1 at $1.45$ dB.

In terms of spectral performance, the AWGs showed good transmission characteristics with 3-dB transmission bandwidths of $56$ pm - $152$ pm for 
TE polarized input signals, corresponding to estimates of spectral resolving powers between $10,200$ and $27,600$. Designs A-1 and D-3 performed close to theoretical 
specifications, reaching $92.5\%$ and $92\%$ of their theoretical resolving power, respectively. Designs A-2, C-4, B-3 and E-5 performed slightly worse 
at $89.4\%$, $87\%$, $86.6\%$ and $86\%$, respectively. Design B-5, while exhibiting the highest resolution, was most severely affected by fabrication errors, 
reaching $79.3\%$ of its theoretical performance. The measured variation of the wavelength-dependent resolving power was found to be $1.7\%$ - $5.8\%$
in the main spectral order and $6\%$ - $15.6\%$ in the entire wavelength range of the measurements. The lowest variation of $R$ was observed
in the designs A-1, A-2, B-3, D-3 and E-5 with $6\%$ - $8.4\%$. In the designs B-5 and C-4, the variation of $R$ was strongest at $14\%$ and $15.6\%$,
respectively.

Measurements of the transmission with TM-polarized input signals showed a polarization dependent wavelength shift $PD\lambda$ between $27$ pm in design A-2 and
$65$ pm in design E-5 despite a square waveguide core profile. 
%The polarization dependence of the AWGs resulted in a degradation of spectral resolving power by $5.9\%$ in design E-5 - $49.3\%$ in design B-5. Design A-2 achieved the highest resolving power in unpolarized mode with $R=19,000$, followed by designs B-3 and C-4 with $R=16,000$, designs A-1 and B-5 with $R=15,000$ and $R=14,000$, respectively. The only designs falling below 10,000 were D-3 and E-5 with $R=9,800$ and $R=9,600$, respectively. 
In a numerical study of the experimental results, we have used a Monte-Carlo algorithm to fit the models of the AWG devices with measured data in order to
calculate the expected spectral image after removal of the output waveguides. Simulations of the fitted AWG models have shown resolving powers of 12,880 (E-5) - 36,000 (B-5)
to be expected in the direct-imaging regime. Furthermore, the obtained model fits were used to study the impact of polarization sensitivity on the resolving power of the AWGs.
The polarization dependence resulted in a degradation of spectral resolving power by $9.4\%$ in design A-1 - $35.2\%$ in design C-4. Design A-2 achieved the highest resolving power in unpolarized mode with $R=24,000$.

We have successfully demonstrated functional custom-tailored AWG designs with an anastigmatic geometry, achieving spectral resolving powers in the near infrared of up 
to $36,000$ in the polarized case and up to $24,000$ in the unpolarized case. 

The new AWG prototypes presented in this work mark an important milestone on the way towards highly integrated, miniaturized, 
medium to high resolution spectrographs for astronomy, that may also find applications in other fields. In view of the saving of 
volume and mass in comparison to bulk optics spectrographs, they are particularly promising for airborne and space applications.
 The anastigmatic spectroscopic AWG devices will be tested on-sky in the PAWS spectrograph \cite{hernandez2020} in the near future, supplanting
the first generation of Rowland-type AWG designs presented in paper I.

\begin{backmatter}
\bmsection{Funding}
The authors gratefully acknowledge support from the BMBF program "Unternehmen Region - Zentren f\"ur Innovationskompetenz"  (grant No. 03Z22A511).

\bmsection{Acknowledgments}
The authors gratefully acknowledge support from the BMBF program "Unternehmen Region - Zentren f\"ur Innovationskompetenz"  (grant No. 03Z22A511).

\bmsection{Disclosures}
The authors declare no conflicts of interest.

\bmsection{Data availability} Data underlying the results presented in this paper are not publicly available at this time but may be obtained from the authors upon reasonable request.

\bmsection{Supplemental document} See Supplement 1 for supporting content.
\end{backmatter}

\bibliography{references}

\end{document}

\maketitle

\section{Lithographic AWG device layouts}
This section contains the lithographic AWG layouts not included in the main document.
\begin{figure}[htbp]
\centering
\fbox{\includegraphics[width=.7\linewidth]{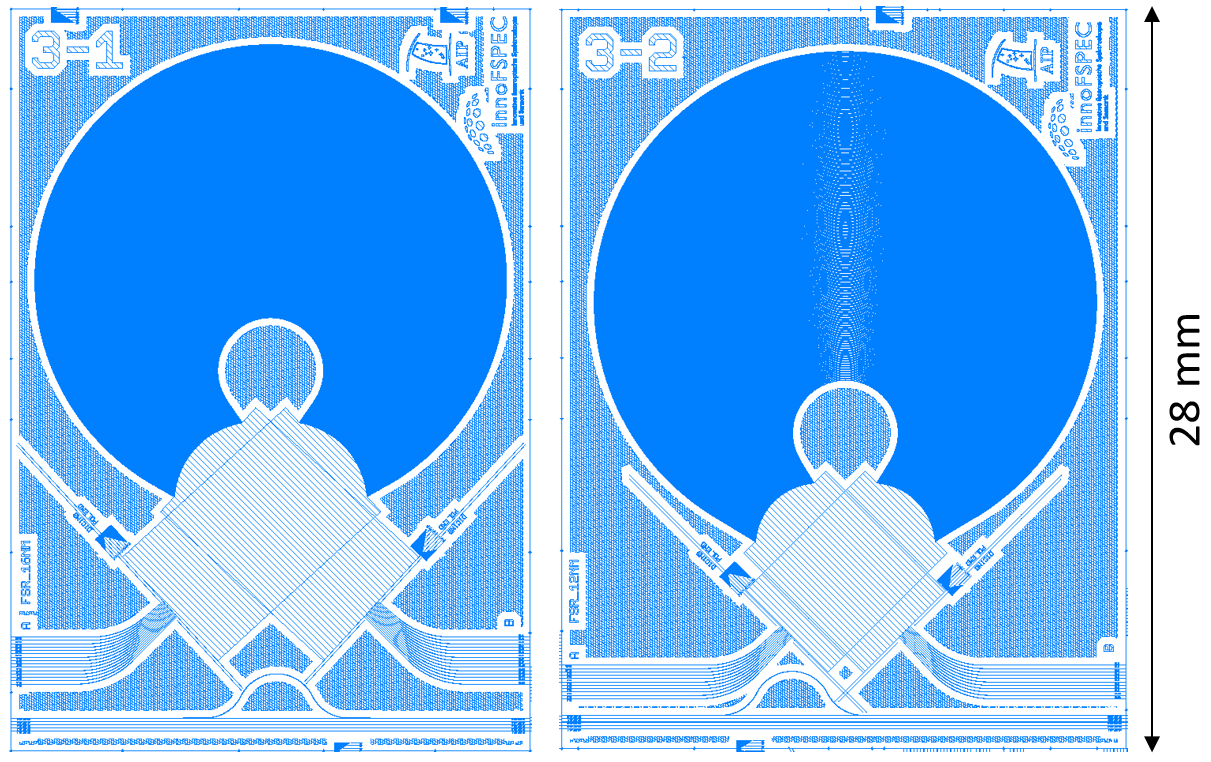}}
\caption{Lithographic mask layouts of designs B-3 (left) and A-2 (right).}
\label{fig:litho_A2_B3}
\end{figure}

\begin{figure}[htbp]
\centering
\fbox{\includegraphics[width=.7\linewidth]{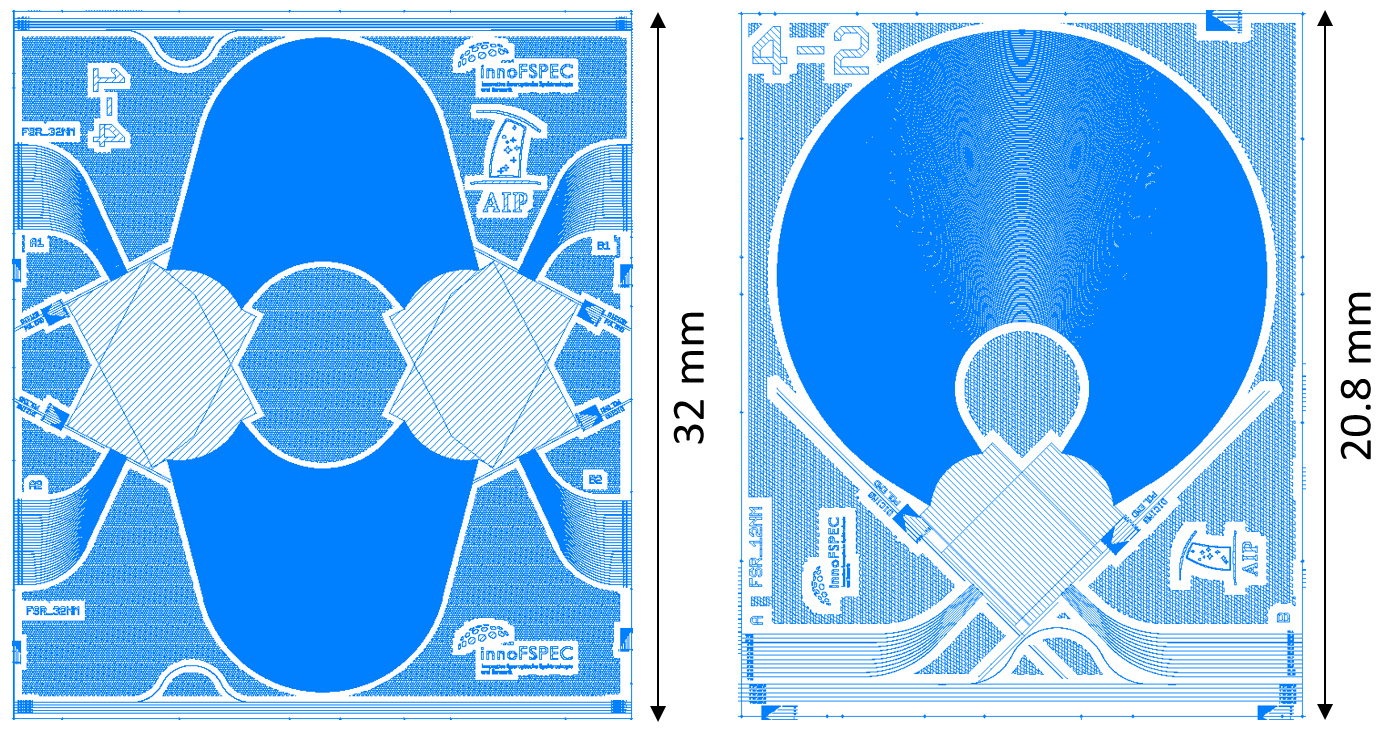}}
\caption{Lithographic mask layouts of designs D-3 (left) and A-1 (right).}
\label{fig:litho_A1_D3}
\end{figure}

\section{Channel transmission curves}
Here, we show the channel power transmission curves omitted in the main document.
\begin{figure}[!htbp]
\centering
\fbox{\includegraphics[width=1.\linewidth]{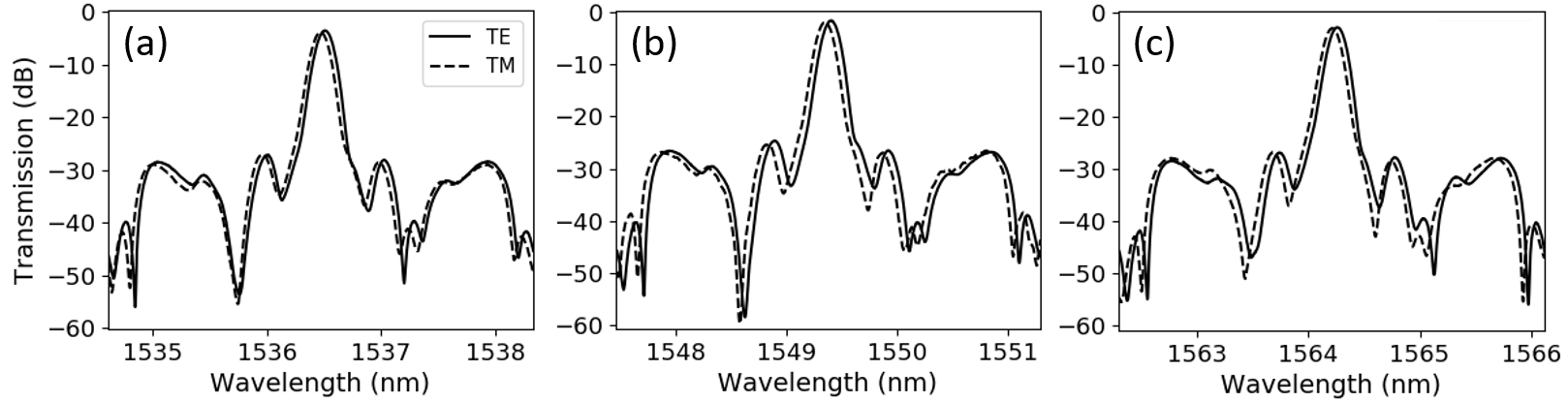}}
\caption{Measured transmission curves of the peripheral channels (a, c) and central channel (b) of device D-3.}
\label{fig:trans_D-3}
\end{figure}

\begin{figure}[!htbp]
\centering
\fbox{\includegraphics[width=1.\linewidth]{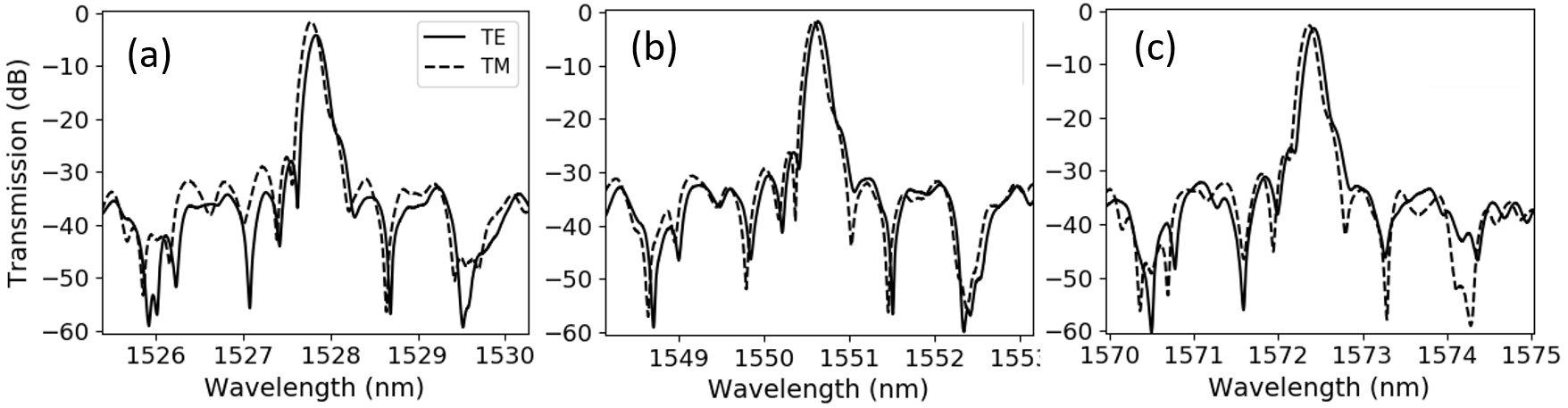}}
\caption{Measured transmission curves of the peripheral channels (a, c) and central channel (b) of device E-5.}
\label{fig:trans_E-5}
\end{figure}

\begin{figure}[!htbp]
\centering
\fbox{\includegraphics[width=1.\linewidth]{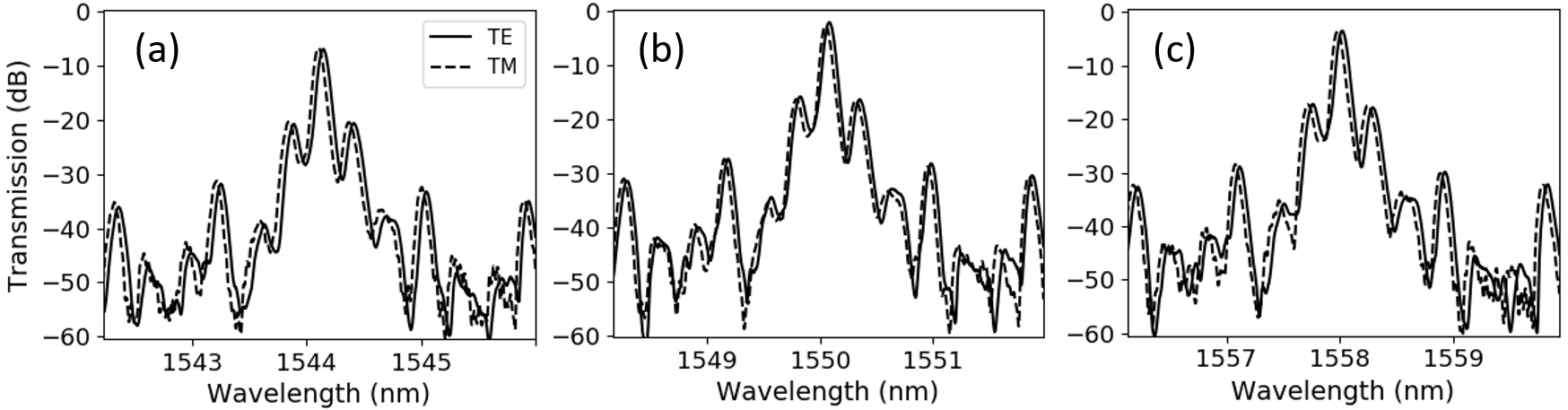}}
\caption{Measured transmission curves of the peripheral channels (a, c) and central channel (b) of device B-3.}
\label{fig:trans_B-3}
\end{figure}

\begin{figure}[!htbp]
\centering
\fbox{\includegraphics[width=1.\linewidth]{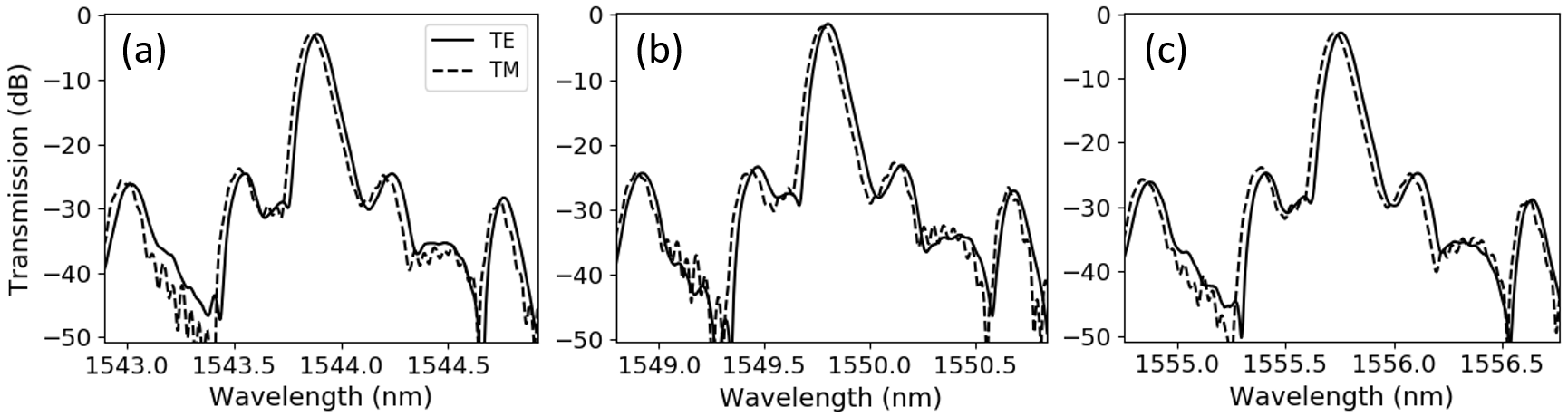}}
\caption{Measured transmission curves of the peripheral channels (a, c) and central channel (b) of device A-1.}
\label{fig:trans_A-1}
\end{figure}